\input{aipcheck}

\documentclass[
    ,final            
  ]
  {aipproc}

\layoutstyle{8x11single}

\begin{document}

\title{Flares in Gamma Ray Bursts}

\classification{98.70.Rz}
\keywords      {$\gamma$-ray sources; $\gamma$-ray burst.}

\author{G. Chincarini}{
  address={INAF, Osservatorio Astronomico di Brera, Via E. Bianchi 46, I-23807, Merate (LC), Italy }
 , altaddress={Universit\'a degli Studi di Milano-Bicocca, Dipartimento di Fisica, Piazza della Scienza 3, I-20126 Milano, Italy }
}

\author{J. Mao}{
address={INAF, Osservatorio Astronomico di Brera, Via E. Bianchi 46, I-23807, Merate (LC), Italy }
}

\author{F. Pasotti}{
address={INAF, Osservatorio Astronomico di Brera, Via E. Bianchi 46, I-23807, Merate (LC), Italy }
 ,altaddress={Universit\'a degli Studi di Milano-Bicocca, Dipartimento di Fisica, Piazza della Scienza 3, I-20126 Milano, Italy }
}

\author{R. Margutti}{
  address={INAF, Osservatorio Astronomico di Brera, Via E. Bianchi 46, I-23807, Merate (LC), Italy }
  ,altaddress={Universit\'a degli Studi di Milano-Bicocca, Dipartimento di Fisica, Piazza della Scienza 3, I-20126 Milano, Italy }
}

\author{C. Guidorzi}{
  address={INAF, Osservatorio Astronomico di Brera, Via E. Bianchi 46, I-23807, Merate (LC), Italy }
}

\author{M. G. Bernardini}{
address={Universit\'a di ``Roma La Sapienza'', ICRA, ICRANet, Italy }
}


\begin{abstract}
The flare activity that is observed in GRBs soon after the prompt emission with the XRT (0.3-10 KeV) instrument 
on board of the Swift satellite is leading to important clues in relation to the physical characteristics of the 
mechanism generating the emission of energy in Gamma Ray Bursts. We will briefly refer to the results obtained with 
the recent analysis [1] and [2] and discuss the preliminary results we obtained with a new larger sample of GRBs 
[limited to early flares] based on fitting of the flares using the Norris 2005 profile. We find, in agreement with 
previous results, that XRT flares follow the main characteristics observed in [3] for the prompt emission spikes. The 
estimate of the flare energy for the subsample with redshift is rather robust and an attempt is made, using the redshisft 
sample, to estimate how the energy emitted in flares depends on time. We used a $H_0=70 km/s/Mpc$, $\Omega_\Lambda=0.7$,
$\Omega_m=0.3$ cosmology. 
\end{abstract}

\maketitle

\section{introduction}
Thanks to the reasonably large amount of data we are obtaining with the Swift satellite [4] we are slowly improving our understanding of the mechanism generating the 
afterglow. Bright bursts as in the case of GRB080319B [5] shed light on the emission mechanisms of the likely multi-component jet and on its time evolution and, however, 
basic uncertainties remain regarding the mechanism of emission and the characteristics of the central object. On the other hand the nature itself of the phenomenon, high 
variability, tell us that the clues of the event are indeed related to the observed variability so that understanding it on various temporal scale may guide toward the 
comprehension of the phenomenon and to the construction of a realistic model.

To this end we decided to approach the problem in a systematic way using different time scales. In addition to the temporal analysis of the prompt emission on the time 
scale of the spikes width and the temporal decay of the afterglow, Pasotti et al. (in preparation), we started the systematic analysis of the flares on temporal scale of 
the flare width [larger than tens of seconds], Chincarini et al. (in preparation), and of the analysis of the prompt emission, afterglow and flares of bright GRBs on 
temporal scales from less than 0.1 s to tens of seconds, Margutti et al., (in preparation). These analyses are based on the data reduction of the 
BAT (Guidorzi et al., (in preparation)) and XRT data carried out by the Italian Swift group, Pasotti et al., (in preparation). In brief while the BAT data have been 
reduced using the standard   time resolution and binning, the XRT data have been both reduced using a standard $S/N$ along the light curve and a binning that optimizes 
the number of photons per bin for each GRB. Spectra, on the other hand, are based on a minimum of 2000 photons have been measured as a function of time and the light 
curve fluxes have been derived from counts using a conversion factor that changes with time according to the measured spectra. The spectra have been measured both 
using a constant mean $N_H$ and leaving $N_H$ as a free parameter. At the time of writing the archives were up dated to April 2008 and the present sample covers the period 
April 2005-April 2008.

\section{The flares sample and flare definition} 
Both in the flare sample and in the flare definition we have a certain degree of subjectivity. The first decision of the analyst is that of fitting the underlying 
light curve. It is assumed, see [1] and [2] and references therein, that the flares are superimposed on an otherwise rather regular afterglow light curve. In most cases
the eye is unambiguously guided in deciding the proper functional shape while in others it remains, due to lack of contiguous observations, open to slightly different 
interpretation.

This leads to uncertainties. The $\chi^2$ value not always allows to distinguishing between different models, however the parameters of the flares generally remain, 
even using different profiles, within the uncertainties of the analysis. Whether a rather faint flare is or not present is again a somewhat subjective decision that 
depends on the observer. This can be checked statistically afterwards but in this sample we limited the selection to flares that were reasonably easily detected by the 
eye; the mini-flare sample is not yet at hand and in any case to the mini flare statistics a high resolution temporal analysis [6] may be preferable. The sample is rather 
sharply defined by the distribution of flare fluence, Figure 1, and flare to underlying light curve continuum ratio (in preparation).

Finally two related questions may create a sample bias as far as early flares are concerned: a) in some cases the early flares, often blended, may be fitted by the 
fitting functions without subtracting the underlying continuum. In this case the decay of the blended flares mimics the underlying continuum. A good example, but not 
the only one, is GRB060714, Figure 2 and Krimm et al., 2007 and b) early flares may simply be the extension of the prompt emission spikes at lower energies especially 
when we are dealing [but often we do not have the redshift] with high redshift objects.

\section{The flare fitting function} 
In [1] we used for the rising part a power law or exponential profile while for the decaying part we used a power law [the fitting of the total observed light curve 
using simultaneously a broken power law and various Gaussian functions was used only for the measure of width and intensity peak since it was shown that such parameters 
were not sensitive, within errors, to the fitting function selected]. The inconvenient of this function is the cusp at the peak that likely does not corresponds to 
reality, furthermore the derivations of fluence lack in part of homogeneity. Similar inconvenient (cusps) are presented by other profile that we 
used as the Norris [8] and Kobayashi\cite{Kobayashi} that were used to fit the prompt emission spikes. Kocevski, Ryde and Liang\cite{Kocevski}  proposed for the prompt 
emission spikes 
a function that fits quite satisfactory the profile of the prompt emission, does not have any cusp and through which the definition of the flare parameters is quite 
robust. Norris 2005 defined another function that also does not have a cusp, fits properly the prompt emission and allows the robust estimate of the spikes parameters. 
This works quite well also for the flares and after experimenting with the various functions proposed in the literature, we decided to choose the latter for simplicity 
and to ease the comparison with the parameters derived by Norris for the prompt emission spikes. This function, as shown in the Figure 3, works very nicely also for 
blended flares. 

\section{Preliminary results} 
The present sample consists of 56 GRBs and for 20 of these we have redshift. Of these 36 GRBs present a single flare, 15 two flares and in 5 cases we were able to 
measure 3 flares. The total number of flares of the sample is therefore 81. Heavily blended multi-component flares are discussed separately to check whether the 
flare parameters are indeed the same as for single flares. 

The flare width is now defined as the distance between the rising and decaying profile at the folding of the peak intensity [width of the profile at I = 0.37 $I_{Peak}$]. 
A rather symmetric profile can be fitted reasonably well also by a Gaussian profile and in this case the width as defined above corresponds 
to $\sigma_{Gauss}$ and practically coincides with the Gaussian width measured at 0.37 IPeak [within about 1$\%$]. This allows
easy comparisons with previous values measured by Chincarini et al. since the two definition of width are related by the relation: $\sigma_{Gauss}\sim w_{Nor}/2.83$. 
In Figure 4 we plot the ratio $Width/T_{peak}$ as a function of $T_peak$, as it can be seen the ratio remains rather constant as a function of time with a 
mean value: $Width/T_{peak}=0.29\pm 0.53$.

The large error is due to the fact, as it can be seen by the large number of points in the figure, that we plotted - to make it easier to give the global view - the 
values derived in all the 5 bands: 0.3 - 10 keV, 0.3 - 1 keV, 1 - 2 keV, 2 - 3 keV and 3 - 10 keV. In some of the energy bands used the counts are smaller and the 
uncertainties larger. Note that according to the above relation this would correspond using a Gaussian fit to $Width/T_{peak}=0.11$ that is in excellent agreement with 
the value measured in [1]: $Width/T_{peak}=0.13\pm 0.10$.

We computed all the parameters derived by Norris et al. 2005 for the prompt emission spikes to verify whether or not we have similar correlations. 
We do (the verious correlations will be given elsewhere). The spikes observed during the prompt emission have a large spread in symmetry for each width. 
Flares observed during the afterglow seem to have a rather constant mean asymmetry $<> = 0.37 \pm 0.58$ and rather large spread, Figure 4.

One of the main reasons why we derived all the light curves in 5 energy channels is to estimate the correlation of the width of the flares with energy. 
These data will be complemented (in progress), when available, with measures of the same flares detected simultaneously by the BAT instrument. Indeed 
due to the much broader energy coverage in this case the result will be more robust. In Figure 5 we show the result of the analysis obtained using the 
XRT data alone from which it is very clearly shown that the median width of the soft energy channel is larger than the median width of the harder energy channel. 

\begin{figure}
  \includegraphics[scale=0.7]{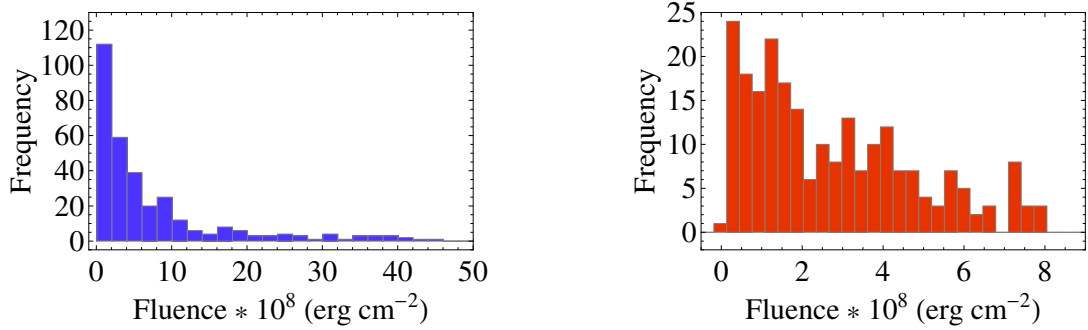}
  \caption{Distribution of the flare fluence measured on the present sample. The histogram, for illustration reasons, contains flare measured in the 
different XRT channels listed in the text. The minimum value measured for the fluence is $8.2\times 10^{-10} cm{-2}$ and the maximum 
$2.1\times 10^{-5}cm^{-2}$. Errors are in the range $1\times 10^{-10} cm^{-2}$ and $2.4\times 10^{-6}cm^{-2}$.}
\end{figure}

\begin{figure}
  \includegraphics[scale=0.7]{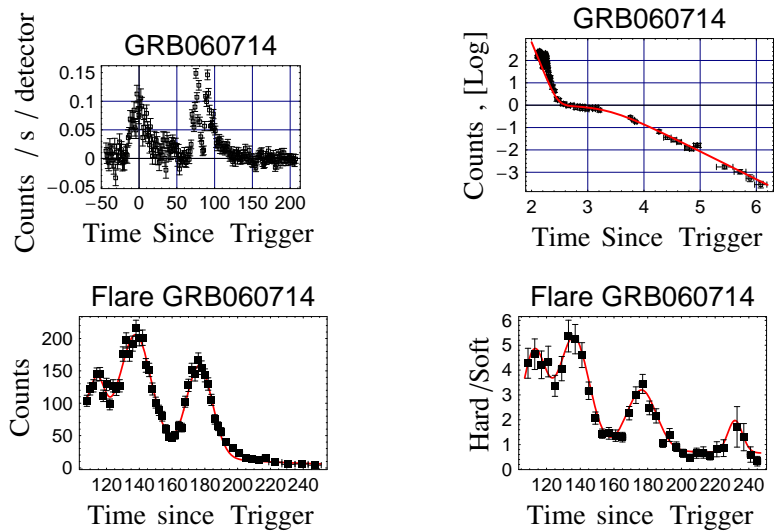}
  \caption{Top left the prompt emission light curve of GRB060714 as detected by the BAT instrument. Top right is used as an example of the fitting of 
the underlying light curve. Note however that the three flares plotted on the bottom left of the figures can also be fitted satisfactory without 
subtracting the underlying light curve shown on top right of the Figure. In other words the (in this case apparent) steep early slope detected by 
XRT may simply not exist and be due to the flares themselves. See [7]  for a detailed discussion of this GRB. On the bottom right the evolution of 
the hardness ratio miming the light curve.}
\end{figure}

\begin{figure}
  \includegraphics[scale=0.8]{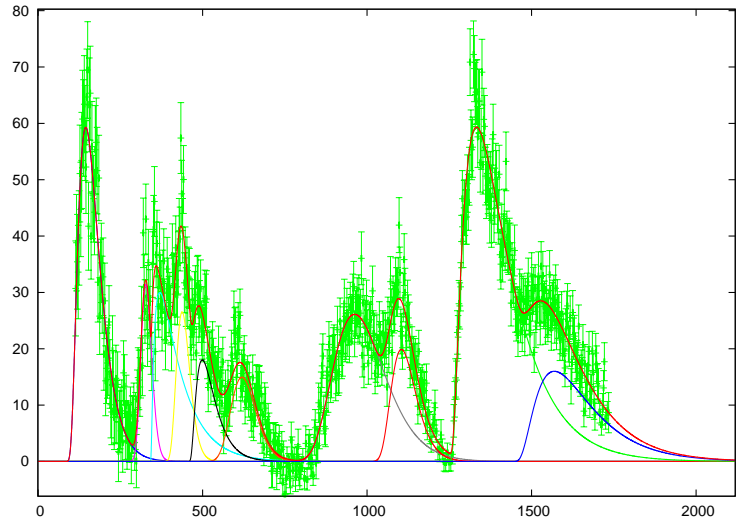}
  \caption{The flare activity in GRB051117A has been fitted, after subtraction of the underlying light curve, with 10 profiles. The fit is excellent 
and shows that after 1500 seconds we have flares that are as bright as those observed at the very beginning of the XRT observations.}
\end{figure}

\begin{figure}
  \includegraphics[scale=0.7]{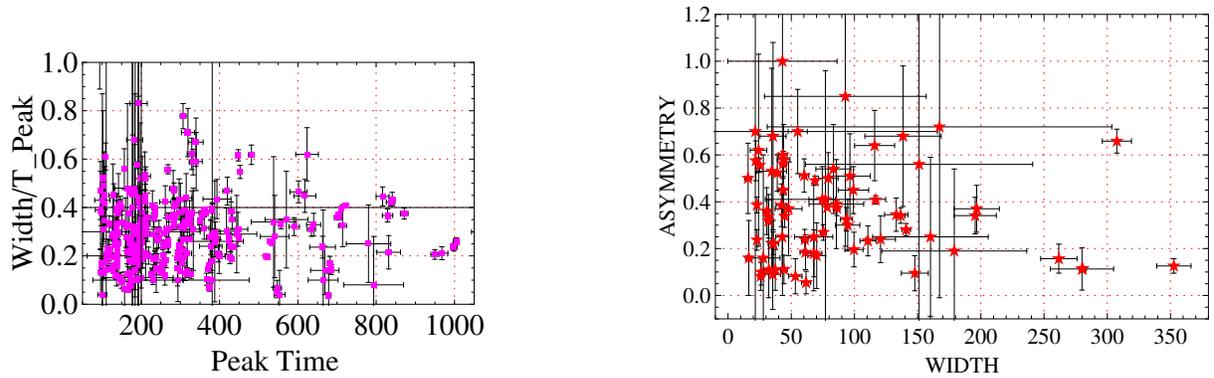}
  \caption{Left: The plot represents all the measures we obtained for the various flares in the five XRT channels defined in the text. This has been done  for 
illustrative purposes and however we increase the scatter since flares in the various channels have obviously less counts that in the 
0.3 - 10 keV total channel. Right: The asymmetry: $\frac{\tau_{decay}-\tau_{rise}}{\tau_{decay}+\tau_{rise}}$  , as a function of the width for 76  flares in the 
channel 0.3 - 10 keV. The mean value is 0.37 with a standard deviation of 0.58.}
\end{figure}


\begin{figure}
  \includegraphics[scale=0.6]{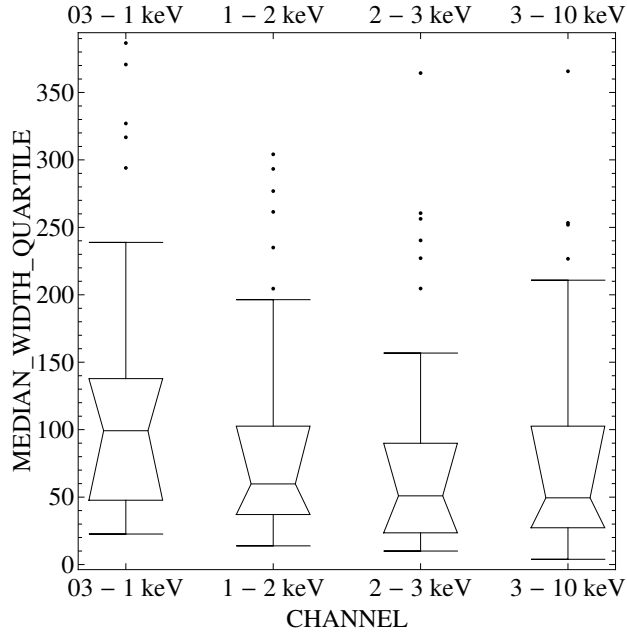}
  \caption{The Box diagram illustrates the width of the flares as a function of the XRT energy channel. The width decreases as a function of energy.}
\end{figure}

\begin{figure}
 \includegraphics[scale=0.6]{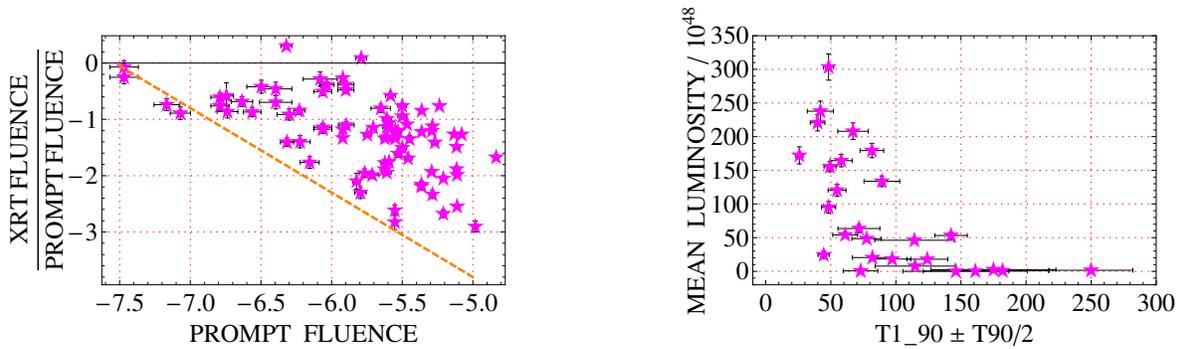}
  \caption{Left: Distribution of the ratio (log) of fluence observed respectively with XRT [channel 0.3 - 10 keV] versus the fluence (log) of the prompt emission. 
A similar plot, as stated in the text, is obtained for the redshift sample where we can transform the fluence in energy [see COSPAR paper]. The dashed line 
has a slope of -1.5 and due very likely to a sensitivity selection bias. Right: The mean luminosity has been computed spreading the flare energy over the flare T90. 
As in Lazzati et al. 2008 this would correspond to a rectangular profile. Note however that this approach does not account for the band correction due to the different 
redshifts and that it largely overestimates the decaying part of the profile so that it should underestimate the slope of the ``global flare light curve''.}
\end{figure}


\section{Energy and Luminosity profile}
Of the 20 GRBs with measured redshift 11 have a single flare, 7 two flares and 2 three measurable early flares for a total of 31 flares. We do not detect any 
correlation of the distribution with redshift [this means we do not have a distance bias] except that at larger z the dispersion in energy is larger with more 
GRBs with high energies [likely a cosmological volume effect].

The flare energetic will further and robustly constrain the models and related efficiency.  If we plot the energy observed in the prompt emission 
[15 - 150 keV] versus the energy observed in Flares we have an (almost) expected result.  GRBs with fainter prompt emission have also a fainter total emission in flares. 
This result is however somewhat weak since we have only two faint prompt emission flares of which on is a short GRB [GRB070124] and the second had a controversial 
redshift estimate [GRB060512 for which we assumed z = 0.4428 \cite{Bloom}]. In the large prompt emission energy region of the plot we have also a tendency of a decreasing 
percentage of the flare emission respect to the energy emitted in the prompt emission. It is however premature to conclude that faint GRBs may have a more efficient 
flare emission since selection effects are not yet fully understood. In Figure 6 we plot the prompt emission fluence versus the ratio XRT/BAT fluence to evidence a 
possible bias due to a) XRT sensitivity limit and b) the sample of faint flares is statistically not significant and it may be improper to mix, to this end, 
short and long GRBs.

Lazzati et al., [12], estimate the average flare light curve of the subsample of Swift afterglows that show flaring activity and for which redshift has been measured. 
While of course this is one of the things that need to be done, we have many doubts on the validity [completeness] of the samples we have at hand at present and on the 
procedures. For instance it is true that for GRB050904 we observe a flare light curve with slope of about -1 but it is also true that this slope is not observed, for 
instance, in GRB071117A and other flares. We will further discuss this point at the COSPAR meeting proceedings and present part of the analysis we are completing on 
this matter. The mean flare luminosity of the GRBs of the present sample and the related T90 is given in Figure 6.





\bibliographystyle{aipproc}   


\begin{thebibliography}{}

\bibitem{Burrows2005}
Chincarini, G., Moretti, A., Romano, P., et al., 2007, Ap.J., 671, 1903

\bibitem{Falcone2006}
Falcone, A.D., Morris, D., Racusin, J., et al., 2007, Ap. J., 671, 1921

\bibitem{Chincarini2007}
Norris, J.P., Bonnell, J.T., Kazanas, D., et al., 2005, Ap.J., 627, 324

\bibitem{Falcone2007}
Gehrels, N., Chincarini, G., Giommi,P., et al., 2004, Ap. J., 611, 1005

\bibitem{Norris2005}
Racusin,J., L., Karpov, S. V., Sokolowski, M., et al., 2008, Nature, submitted

\bibitem{Butler2007}
Margutti, R., These proceedings.

\bibitem{Lazzati2007}
Krimm, H.A., Granot, J., and marshall, F.E., et al.,  2007 ApJ 665, 554

\bibitem{Kumar2008}
Norris, J.P., Nemiroff, R.J., Bonnell, J.,T., et al., 1996, Ap.J. 459, 393

\bibitem{Kobayashi}
Kobayashi, S., Piran, T., and Sari, R., 1997, Ap.J., 490, 92

\bibitem{Kocevski}
Kocevski, D., Ryde, F., and Liang, E., 2003, Ap.J., 596, 389

\bibitem{Bloom}
Bloom, J.,S., Foley, R.J., Kocevski, D., Perley, D., 2006, GCN 5217

\bibitem{Curran2008}
Lazzati, D., Perna, R., and Begelman, M.C., 2008, MNRAS


\end{thebibliography}




\begin{theacknowledgments}
The GRB Research is being supported by ASI contract SWIFT I/011/07/0 and by MIUR PRIN 2007TNYZXL. The continuous support by the Universita degli Studi di Milano 
Bicocca is also acknowledged.
\end{theacknowledgments}

\bibliographystyle{aipproc}   

\end{document}